\documentclass{vgtc}                          

\graphicspath{{figures/}{pictures/}{images/}{./}} 
\usepackage{times}                     

\usepackage{tabu}                      
\usepackage{booktabs}                  
\usepackage{lipsum}                    
\usepackage{mwe}                       

\usepackage{mathptmx}                  

\usepackage{enumitem}

\usepackage{pdflscape}
\usepackage{diagbox} 

\onlineid{0}

\vgtccategory{Research}

\vgtcinsertpkg

\title{A Design Space of Visual Interfaces for Generative Image Models}

\author{Susie S.Y. Li \thanks{e-mail: suyang.li@tufts.edu}\\ %
        \scriptsize Tufts University %
\and Mingwei S.G. Li\\ %
     \scriptsize Tufts University %
\and Remco Chang\\ %
     \scriptsize \centering Tufts University
}

\teaser{
  \centering
  \includegraphics[width=0.85\linewidth]{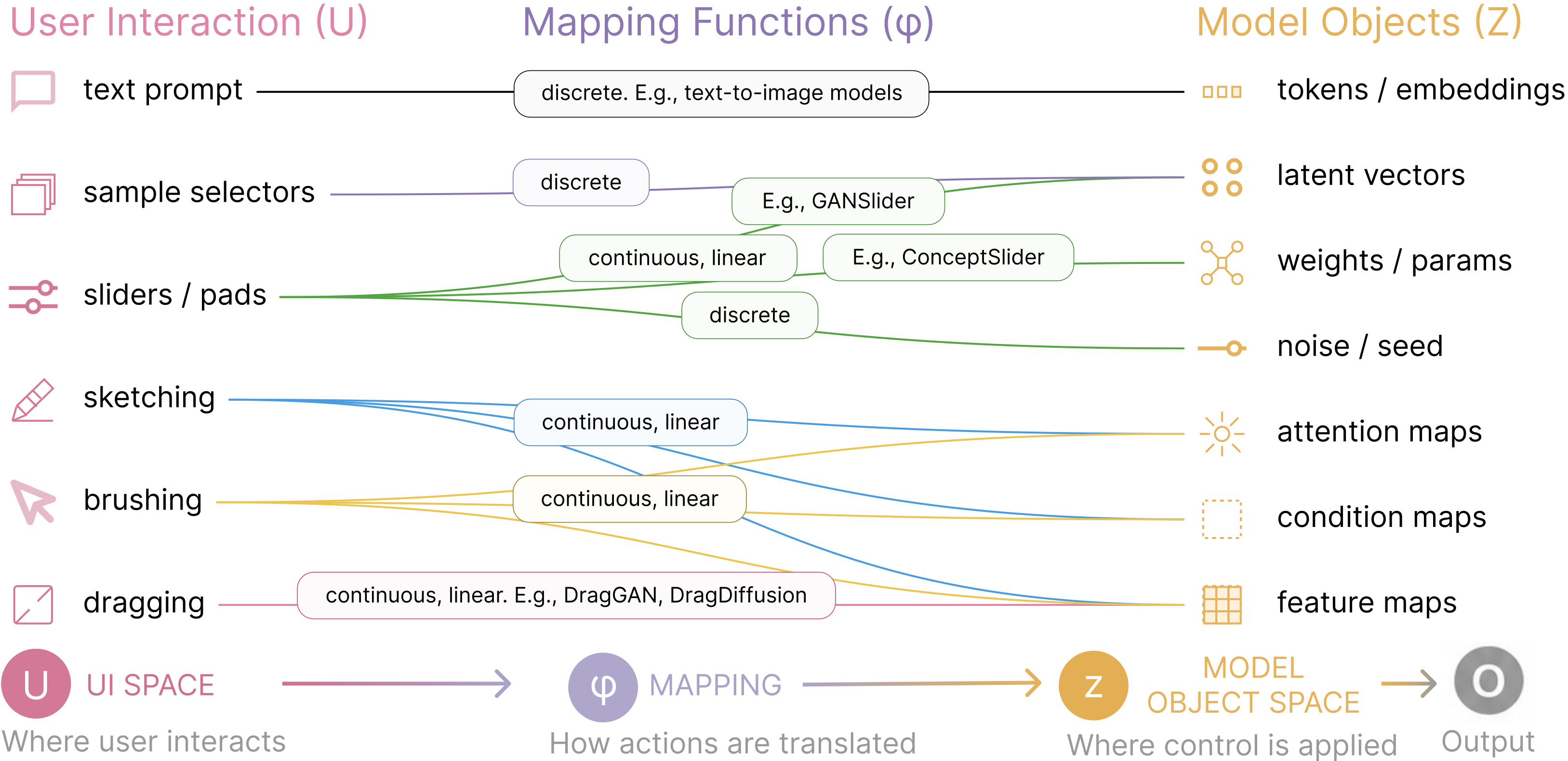}
  \caption{The framework introduces three complementary spaces: User Interface, Controlled Model Objects, and Mapping Functions between them.}
  \label{fig:teaser}
}

\abstract{
Interactive visual interfaces have become an important means of controlling generative image models, enabling users to manipulate generation through prompts, direct manipulation, and a range of interactions. However, existing techniques are typically presented as independent systems, making it difficult to understand how they relate, compare their interaction mechanisms, or identify opportunities for new interface designs. We introduce a design space for interactive visual interfaces for generative image models derived from 51 research systems and practitioner tools. The framework decomposes each system into three complementary components: the user interface (U), the controllable model objects (Z), and the mapping function ($\phi$) that translates user interaction into model operations. This decomposition provides a common representation for analyzing heterogeneous interaction techniques across model families, revealing recurring design patterns and underexplored regions of the design space. We further present an interactive corpus explorer that support comparative analysis, and discuss usage scenarios for both educational settings and HCI/AI practitioners identifying research and design opportunities.
}

\keywords{Generative models, interactive visual interfaces}

\begin{document}


\firstsection{Introduction}

\maketitle


Recent advances in generative AI have been accompanied by a growing diversity of interactive visual interfaces that enable users to explore, edit, and steer image generation. Existing systems expose a wide range of interaction techniques, including text prompting~\cite{ hertz2022prompttopromptimageeditingcross}, semantic sliders~\cite{GANSlider, gandikota2025sliderspace}, sketching~\cite{nazeri2019edgeconnect}, dragging~\cite{Pan2023draggan, shi2024dragdiffusion}, brushing~\cite{Chung_2023PromptPaint}, inpainting~\cite{bau2019gandissection}, and example-based editing~\cite{Evirgen_2022_GANzilla, wan_gancollage_2023}, each providing different mechanisms for manipulating the underlying generative model. 

Despite this growing body of work, existing interaction techniques are typically described in terms of either their visible interface or their underlying model architecture, making it difficult to systematically compare systems or understand how user interactions are translated into model behavior. To address this gap, we curated and analyzed a corpus of 51 interactive generative image system spanning research prototypes and practitioner tools. Analysis of this corpus revealed that existing systems can be understood through three complementary components: \textbf{user interface (U), controllable model objects (Z),} and the \textbf{mapping function ($\phi$)} between them (Figure~\ref{fig:teaser}). Building on this observation, we propose a unified design space that characterizes interactive interfaces for controlling generative image models through these three complementary dimensions. To demonstrate its utility, we develop an interactive corpus explorer that supports systematic comparison of existing techniques and helps students, HCI researchers, and AI practitioners analyze how interaction designs relate to underlying model mechanisms. 

\section{Corpus Collection and Analysis}

We constructed and coded the corpus using the three-stage Initialize–Expand–Refine methodology of Solen et al.\cite{solen2025designspacemultiscale}, in which corpus development and framework refinement proceed iteratively. We began with 21 author-known academic and practitioner systems, then expanded the corpus using four recent surveys as structured entry points, adding 33 systems from major HCI, visualization, and machine learning venues such as CHI, VIS, UIST, NeurIPS, ICML, and ICLR. After screening against explicit criteria, three systems were removed, yielding a final corpus of 51. Our primary scope covers work from 2021–2025, while retaining earlier  work when needed for historical context. Included systems had to introduce a novel interaction technique for generative control, support iterative modification under explicit user intent, and alter or navigate the model’s learned generative distribution through mechanisms such as latent traversal, conditioning, feature modulation, or parameter adaptation. We excluded systems with only trivial controls, open-ended ideation or browsing without explicit steering, and purely analytic tools that did not modify the generative process (e.g.\cite{Jeong_2023}). The final corpus includes both academic and practitioner-facing systems to capture the broader landscape of interactive generative control.


\section{Design Space}
The proposed design space decomposes interactive generative interfaces into three complementary components: the \textbf{UI Space}, \textbf{Model Object Space}, and \textbf{Mapping Space}. Rather than treating an interaction technique as a single system, this decomposition separates how users express intent, what is controlled inside the generative model, and how these two spaces are connected. This enables interaction techniques from different model families and interface paradigms to be compared using a common representation.

\smallskip
\noindent\textbf{UI Space} 
captures the interaction mechanism visible to users. We identify three broad categories of interaction primitives:\\
1) \textit{Direct input conditions} include text prompts, reference images, and spatial conditioning inputs such as masks or layouts;\\
2) \textit{Parameterized controls} include sliders and 2D control pads that expose continuous navigation through latent space;\\
3) \textit{Direct manipulation} includes dragging, brushing, inpainting, and outpainting, where users edit image regions directly. 

\smallskip
\noindent\textbf{Model Object Space}
 describes what internal representations are manipulated by an interface. Existing systems expose a wide range of controllable objects, including text embeddings, image embeddings, spatial condition maps, latent vectors, intermediate feature maps, activation tensors, and model weights. These representations differ substantially in their locality, semantic interpretability, persistence, and scale of influence over the generated output. 

\smallskip
\noindent\textbf{Mapping Space}
 defines how interface actions are translated into operations on the underlying model. Existing systems realize this mapping through mechanisms such as latent-space traversal, activation editing, optimization-based inversion, feature modulation, or weight adaptation. We characterize these mappings using several structural properties, including cardinality, linearity, locality preservation, constraint structure, adaptivity, and sensitivity.

 

\begin{table*}[t]
\caption{    Sample works from the survey highlighting key aspects of our design framework}
  \label{table:survey-sample}
  
\begin{tabular}{llllll}
Work                                                                                   & generative model & User Interface (U) & mapping function ($\phi$)                                            & model object (Z)             &  \\
\hline
GANSlider~\cite{GANSlider}                                       & StyleGAN2        & sliders            & $ \;\; w_1=w_0+s\,\Delta w_i, \Delta w_i=\mathrm{PCA}_i(w)$          & style vectors                &  \\
SliderSpace~\cite{gandikota2025sliderspace}                      & Diffusion        & sliders            & $ w_1=w_0+s\,\Delta w_{\mathrm{LoRA}}(c)$                          & model weights                &  \\
Prompt-to-Prompt~\cite{hertz2022prompttopromptimageeditingcross} & Diffusion        & prompt             & $attn = (t>=\tau)?\; attn_{src} \,:\, attn_{new} $ & cross-attention feature maps & 

\end{tabular}
\end{table*}

\section{Interactive Survey Explorer}
To operationalize the design space, we developed an interactive survey explorer that visualizes the coded corpus\footnote{\url{https://susiesyli.com/interactive-survey-explorer/}}. Users can filter systems by interaction technique, model family, controlled model objects, or mapping mechanism, inspect representative systems and techniques, and compare interaction designs through coordinated visualizations. This open source explorer enables interactive analysis of recurring patterns and supports intuitive explorations of underrepresented regions of the design space. 

Leveraging the explorer, we interactively inspect the coded corpus and identify recurring patterns. In particular, we observe that 1) user interaction is primarily realized through text prompts and image-space conditioning, 2) feature maps and low-rank adaptations (LoRA)~\cite{hu2021lora} are the most frequently exposed model objects, and 3) most system employ static one-to-one mappings between interface actions and model operations.  





\section{Usage Scenarios / Design Opportunities}
\subsection{Educational}
Students learning about interactive generative AI are often introduced to systems such as DragonDiffusion\cite{mou2023dragondiffusion}, ControlNet\cite{zhang2023controlnet}, and Prompt-to-Prompt\cite{hertz2022prompttopromptimageeditingcross} as collections of independent techniques. While they may recognize differences in the visible interface, it is much harder to identify whether two systems' behavior and function differ because they expose different model representations, employ different mappings between user input and model behavior, or simply represent the same underlying mechanism through different interface widgets. The proposed framework makes these distinctions explicit by decomposing each system into the three components. This decomposition enables students to isolate where systems differ, compare interaction techniques across model families using a common representation, and reason about how individual design decisions influence controllability and user experience. 

\subsection{Design Implications for HCI/AI Practitioners}
Through the lens of our design framework, the mapping between the UI space (U) and model object space (Z) reveals two research opportunities. First, altering the UI while preserving the model object motivates interfaces better aligned with the underlying representation, such as painting directly on diffusion feature maps instead of editing text prompts. Second, rethinking the mapping itself enables perceptually or task-aware mappings, rather than linear or uniform ones in model object space, yielding more intuitive and predictable interfaces.

\smallskip
\noindent\textbf{Painting on feature maps in diffusion models}: 

A key limitation of current text-to-image systems is the lack of direct, fine-grained spatial control. Methods such as prompt-to-prompt \cite{hertz2022prompttopromptimageeditingcross} manipulate intermediate feature maps (e.g., cross-attention) only indirectly through prompt editing, creating a mismatch between the interaction space (U) and the underlying model object space (Z). This gap motivates feature map painting, where the interaction is painting and the model object is the diffusion model’s intermediate feature map. Instead of repeatedly refining prompts, users directly paint semantic activations to strengthen or suppress concepts in selected regions. Prompt embeddings continue to specify global semantics, while painted feature maps impose explicit spatial constraints during denoising, yielding precise, continuous, and localized control beyond what prompt-based interaction alone affords.

\smallskip
\noindent\textbf{Perceptually uniform slider control}:
GANSlider~\cite{GANSlider} linearly maps slider inputs to StyleGAN2 style directions defined by PCA, while SliderSpace~\cite{gandikota2025sliderspace, gandikota2023conceptsliders} linearly maps sliders to model weight directions via low rank adaptations (LoRA)~\cite{hu2021lora}. Although they manipulate different model objects, both assume a uniform mapping in the model object space, where equal slider increments correspond to equal latent or parameter displacements. This reveals a broader research gap in the properties of the mapping function ($\phi$) rather than the choice of model object. Because the geometry of latent and weight spaces is generally misaligned with human perception, equal distances in these spaces often produce highly non-uniform perceptual changes. Instead, one can learn a non-uniform mapping that compensates for this nonlinear geometry, so that equal slider increments correspond to approximately equal perceptual changes in an embedding space (e.g., CLIP~\cite{radford2021clip} or LPIPS~\cite{zhang2018lpips}). Rather than enforcing uniformity where it is mathematically convenient (the model object space), this approach enforces uniformity where it matters to users, that is, the perceptual space, yielding smoother, more predictable interaction.

\section{Conclusion}
We present a design space for interactive visual interfaces for generative image models derived from a corpus of 51 systems. By decomposing interaction into user interfaces, model objects, and mapping functions, the framework supports systematic comparison and discovery of new interaction techniques. We hope this work provides a common foundation for future research on controllable generative AI interfaces. 









\bibliographystyle{abbrv-doi}
\bibliography{template}

@inproceedings{Jeong_2023,
   title={Concept Lens: Visually Analyzing the Consistency of Semantic Manipulation in GANs},
   url={http://dx.doi.org/10.1109/VIS54172.2023.00053},
   DOI={10.1109/vis54172.2023.00053},
   booktitle={2023 IEEE Visualization and Visual Analytics (VIS)},
   publisher={IEEE},
   author={Jeong, Sangwon and Li, Mingwei and Berger, Matthew and Liu, Shusen},
   year={2023},
   month=Oct, pages={221–225} 
}

@inproceedings{GANSlider, series={CHI '22},
   title={GANSlider: How Users Control Generative Models for Images using Multiple Sliders with and without Feedforward Information},
   url={http://dx.doi.org/10.1145/3491102.3502141},
   DOI={10.1145/3491102.3502141},
   booktitle={CHI Conference on Human Factors in Computing Systems},
   publisher={ACM},
   author={Dang, Hai and Mecke, Lukas and Buschek, Daniel},
   year={2022},
   month=apr, pages={1–15},
   collection={CHI '22} }

@inproceedings{bau2019gandissection,
  title={GAN Dissection: Visualizing and Understanding Generative Adversarial Networks},
  author={Bau, David and Zhu, Jun-Yan and Strobelt, Henning and Lapedriza, Agata and Zhou, Bolei and Torralba, Antonio},
  booktitle={Proceedings of the International Conference on Learning Representations (ICLR)},
  year={2019}
}

@misc{hertz2022prompttopromptimageeditingcross,
      title={Prompt-to-Prompt Image Editing with Cross Attention Control}, 
      author={Amir Hertz and Ron Mokady and Jay Tenenbaum and Kfir Aberman and Yael Pritch and Daniel Cohen-Or},
      year={2022},
      eprint={2208.01626},
      archivePrefix={arXiv},
      primaryClass={cs.CV},
      url={https://arxiv.org/abs/2208.01626}, 
}

@inproceedings{Evirgen_2022_GANzilla, series={UIST '22},
   title={GANzilla: User-Driven Direction Discovery in Generative Adversarial Networks},
   url={http://dx.doi.org/10.1145/3526113.3545638},
   DOI={10.1145/3526113.3545638},
   booktitle={Proceedings of the 35th Annual ACM Symposium on User Interface Software and Technology},
   publisher={ACM},
   author={Evirgen, Noyan and Chen, Xiang "Anthony"},
   year={2022},
   month=oct, pages={1–10},
   collection={UIST '22} }

@inproceedings{wan_gancollage_2023,
	address = {Pittsburgh PA USA},
	title = {{GANCollage}: {A} {GAN}-{Driven} {Digital} {Mood} {Board} to {Facilitate} {Ideation} in {Creativity} {Support}},
	isbn = {978-1-4503-9893-0},
	shorttitle = {{GANCollage}},
	url = {https://dl.acm.org/doi/10.1145/3563657.3596072},
	doi = {10.1145/3563657.3596072},
	language = {en},
	urldate = {2026-01-07},
	booktitle = {Proceedings of the 2023 {ACM} {Designing} {Interactive} {Systems} {Conference}},
	publisher = {ACM},
	author = {Wan, Qian and Lu, Zhicong},
	month = jul,
	year = {2023},
	pages = {136--146},
}

@misc{gandikota2025sliderspace,
      title={SliderSpace: Decomposing the Visual Capabilities of Diffusion Models}, 
      author={Rohit Gandikota and Zongze Wu and Richard Zhang and David Bau and Eli Shechtman and Nick Kolkin},
      year={2025},
      eprint={2502.01639},
      archivePrefix={arXiv},
      primaryClass={cs.CV},
      url={https://arxiv.org/abs/2502.01639}, 
}

@misc{gandikota2023conceptsliders,
      title={Concept Sliders: LoRA Adaptors for Precise Control in Diffusion Models}, 
      author={Rohit Gandikota and Joanna Materzynska and Tingrui Zhou and Antonio Torralba and David Bau},
      year={2023},
      eprint={2311.12092},
      archivePrefix={arXiv},
      primaryClass={cs.CV},
      url={https://arxiv.org/abs/2311.12092}, 
}

@inproceedings{Pan2023draggan, series={SIGGRAPH '23},
   title={Drag Your GAN: Interactive Point-based Manipulation on the Generative Image Manifold},
   url={http://dx.doi.org/10.1145/3588432.3591500},
   DOI={10.1145/3588432.3591500},
   booktitle={Special Interest Group on Computer Graphics and Interactive Techniques Conference Conference Proceedings},
   publisher={ACM},
   author={Pan, Xingang and Tewari, Ayush and Leimkühler, Thomas and Liu, Lingjie and Meka, Abhimitra and Theobalt, Christian},
   year={2023},
   month=jul, pages={1–11},
   collection={SIGGRAPH '23} }

@inproceedings{Chung_2023PromptPaint, series={UIST '23},
   title={PromptPaint: Steering Text-to-Image Generation Through Paint Medium-like Interactions},
   url={http://dx.doi.org/10.1145/3586183.3606777},
   DOI={10.1145/3586183.3606777},
   booktitle={Proceedings of the 36th Annual ACM Symposium on User Interface Software and Technology},
   publisher={ACM},
   author={Chung, John Joon Young and Adar, Eytan},
   year={2023},
   month=oct, pages={1–17},
   collection={UIST '23} }

@misc{nazeri2019edgeconnect,
      title={EdgeConnect: Generative Image Inpainting with Adversarial Edge Learning}, 
      author={Kamyar Nazeri and Eric Ng and Tony Joseph and Faisal Z. Qureshi and Mehran Ebrahimi},
      year={2019},
      eprint={1901.00212},
      archivePrefix={arXiv},
      primaryClass={cs.CV},
      url={https://arxiv.org/abs/1901.00212}, 
}

@misc{zhang2023controlnet,
      title={Adding Conditional Control to Text-to-Image Diffusion Models}, 
      author={Lvmin Zhang and Anyi Rao and Maneesh Agrawala},
      year={2023},
      eprint={2302.05543},
      archivePrefix={arXiv},
      primaryClass={cs.CV},
      url={https://arxiv.org/abs/2302.05543}, 
}

@misc{hu2021lora,
      title={LoRA: Low-Rank Adaptation of Large Language Models}, 
      author={Edward J. Hu and Yelong Shen and Phillip Wallis and Zeyuan Allen-Zhu and Yuanzhi Li and Shean Wang and Lu Wang and Weizhu Chen},
      year={2021},
      eprint={2106.09685},
      archivePrefix={arXiv},
      primaryClass={cs.CL},
      url={https://arxiv.org/abs/2106.09685}, 
}

@misc{solen2025designspacemultiscale,
      title={A Design Space for Multiscale Visualization}, 
      author={Mara Solen and Matt Oddo and Tamara Munzner},
      year={2025},
      eprint={2404.01485},
      archivePrefix={arXiv},
      primaryClass={cs.HC},
      url={https://arxiv.org/abs/2404.01485}, 
}

@misc{shi2024dragdiffusion,
      title={DragDiffusion: Harnessing Diffusion Models for Interactive Point-based Image Editing}, 
      author={Yujun Shi and Chuhui Xue and Jun Hao Liew and Jiachun Pan and Hanshu Yan and Wenqing Zhang and Vincent Y. F. Tan and Song Bai},
      year={2024},
      eprint={2306.14435},
      archivePrefix={arXiv},
      primaryClass={cs.CV},
      url={https://arxiv.org/abs/2306.14435}, 
}

@misc{mou2023dragondiffusion,
      title={DragonDiffusion: Enabling Drag-style Manipulation on Diffusion Models}, 
      author={Chong Mou and Xintao Wang and Jiechong Song and Ying Shan and Jian Zhang},
      year={2023},
      eprint={2307.02421},
      archivePrefix={arXiv},
      primaryClass={cs.CV},
      url={https://arxiv.org/abs/2307.02421}, 
}

@inproceedings{radford2021clip,
  title = {Learning transferable visual models from natural language supervision},
  author = {Radford, Alec and Kim, Jong Wook and Hallacy, Chris and Ramesh, Aditya and Goh, Gabriel and Agarwal, Sandhini and Sastry, Girish and Askell, Amanda and Mishkin, Pamela and Clark, Jack and others},
  booktitle = {International conference on machine learning},
  pages = {8748--8763},
  year = {2021},
  organization = {PmLR},
}

@inproceedings{zhang2018lpips,
  title = {The unreasonable effectiveness of deep features as a perceptual metric},
  author = {Zhang, Richard and Isola, Phillip and Efros, Alexei A and Shechtman, Eli and Wang, Oliver},
  booktitle = {2018 IEEE/CVF conference on computer vision and pattern recognition},
  pages = {586--595},
  year = {2018},
  organization = {IEEE},
}

\end{document}